\documentclass[]{aastex}
\usepackage{emulateapj5,pstricks,psfig}

\newcommand{\beqa}{\begin{eqnarray}}
\newcommand{\eeqa}{\end{eqnarray}}

\def\LCDM{$\Lambda$CDM\/}
\def\LWDM{$\Lambda$WDM\/}
\def\LWDMth{$\Lambda$WDM$_{\rm th}$\/}
\def\hkpc{\ h^{-1}{\rm kpc}}
\def\hMsun{\ h^{-1}M_{\odot}}
\def\hMpc{\ h^{-1}{\rm Mpc}}
\def\gsim {\lower .1ex\hbox{\rlap{\raise .6ex\hbox{\hskip .3ex
        {\ifmmode{\scriptscriptstyle >}\else
                {$\scriptscriptstyle >$}\fi}}}
        \kern -.4ex{\ifmmode{\scriptscriptstyle \sim}\else
                {$\scriptscriptstyle\sim$}\fi}}}
\def\lsim {\lower .1ex\hbox{\rlap{\raise .6ex\hbox{\hskip .3ex
        {\ifmmode{\scriptscriptstyle <}\else
                {$\scriptscriptstyle <$}\fi}}}
        \kern -.4ex{\ifmmode{\scriptscriptstyle \sim}\else
                {$\scriptscriptstyle\sim$}\fi}}}
\def\beq{\begin{equation}}
\def\eeq{\end{equation}}

\begin{document}
\slugcomment{{\em Astrophysical Journal Letters, submitted}}
\lefthead{Angular momentum profiles of WDM halos}
\righthead{BULLOCK ET AL.}

\title{Angular momentum profiles of warm dark matter halos}\vspace{3mm}

\author{James S. Bullock\altaffilmark{1}, Andrey V. Kravtsov\altaffilmark{1,3}}
\affil{Department of Astronomy, The Ohio State University,
    140 W. 18th Ave, Columbus, OH 43210-1173}
\author{Pedro Col\'{\i}n\altaffilmark{2}}
\affil{Instituto de Astronom\'{\i}a, U.N.A.M., A.P. 70-264, 
04510, M\'exico, D.F., M'exico}

\altaffiltext{1}{james,andrey@astronomy.ohio-state.edu}
\altaffiltext{2}{colin@astroscu.unam.mx}
\altaffiltext{3}{Hubble Fellow. Current address: Department
of Astronomy and Astrophysics, University of Chicago,  5640 S. Ellis Ave.,
Chicago, IL 60637}

\begin{abstract}
  We compare the specific angular momentum profiles of virialized dark
  halos  in cold dark matter  (CDM) and warm  dark matter (WDM) models
  using   high-resolution dissipationless simulations. The simulations
  were  initialized   using the same  set  of  modes,  except on small
  scales, where  the  power was  suppressed in WDM  below the filtering
  length.  Remarkably, WDM as well as CDM  halos are well-described by
  the  two-parameter    angular momentum profile  of    Bullock et al.
  (2001), even though the halo masses are below the filtering scale of
  the    WDM.   Although   the   best-fit  shape  parameters    change
  quantitatively for individual halos  in the two simulations, we find
  no {\em systematic} variation in profile shapes as a function of the
  dark matter type.  The  scatter in shape parameters is significantly
  smaller for the    WDM halos, suggesting   that  substructure and/or
  merging   history plays a role    producing  scatter about the  mean
  angular  momentum  distribution, but that  the average angular
  momentum profiles of halos  originate from larger-scale phenomena or
  a mechanism associated with  the  virialization process.  The  known
  mismatch between  the angular momentum  distributions  of dark halos
  and disk galaxies is therefore present in WDM as well as CDM models.
  Our WDM halos tend to have a less coherent (more misaligned) angular
  momentum  structure and smaller spin  parameters  than do their CDM
  counterparts,  although we caution  that this result is  based on a
  small number of halos.
\end{abstract}
\keywords{cosmology: theory -- galaxies:formation}

\section{Introduction}
\label{sec:intro}

Cold Dark Matter (CDM) models  of structure  formation have been  very
successful in describing the observed properties of galaxies and their
spatial   distribution at  large    and intermediate scales  ($\gtrsim
1\hMpc$). At small ($\lesssim 100\hkpc$)   scales, however, there  are
some  uncomfortable discrepancies     between   the observations   and
straightforward CDM  predictions.  One  potential problem concerns the
sizes  and  angular momentum  content   of  disk  galaxies.   Analytic
investigations based within the   CDM framework have   been reasonably
successful in reproducing  the observed  sizes  of disk  galaxies, but
only under the simplistic assumption  that disks  arise from gas  with
the  same  initial  average specific angular  momentum   as their host
halos,  and that  the  gas experiences   little angular momentum  loss
during the   formation  process  \citep[e.g.,][]{fe80,  blum86, dss97,
mmw98,   vdb00,fa00}.   However,  in  more  detailed  numerical galaxy
formation simulations,  the gas appears to loose   a large fraction of
its  initial    angular   momentum    \citep[e.g.,][]{navarro_white94,
navarro_steinmetz00}.   The  resultant  disks are considerably smaller
than observed disks, unless gas cooling is delayed \citep{weil_etal98}
or the efficiency of stellar feedback is enhanced
\citep{thacker_couchman01}.

It  is not yet clear whether  this discrepancy poses a serious problem
for CDM or  is simply  a result  of our  insufficient understanding or
inadequate numerical  modeling of the  complicated processes operating
during galaxy formation.  We can gain  some insight into this question
by comparing the specific angular momentum  ($j$) distribution of dark
matter in galactic halos  to that of  observed for galactic disks.  In
Bullock et al. (2001; hereafter B01), using a dissipationless CDM plus
cosmological constant ($\Lambda$CDM) simulation, we found that angular
momentum profiles of  galactic CDM halos  of mass $M_{\rm v}$ are well
described by a two-parameter angular momentum profile of the form
\begin{equation} 
M(<j) = M_{\rm v} \frac{\mu j}{j_0 + j}, \quad \mu > 1. 
\label{eq:fit} 
\end{equation} 
Here,  $j$ is  projected  along the   direction of  the total angular
momentum in the halo.  The parameters $\mu$ and $j_0$ fully define the
angular  momentum content of  the halo,  where  the global spin of the
halo\footnote{This  spin parameter is a  practical modification of the
conventional spin  parameter, defined  as $\lambda=J\sqrt{|E|}/GM_{\rm
v}^{5/2}$, where $E$  is the halo  internal  energy.} $\lambda' \equiv
J/\sqrt{2}M_{\rm v}V_{\rm v}R_{\rm v}$, is  related to $\mu$ and $j_0$
via $ j_0\, b(\mu)  = \sqrt{2} V_{\rm  v} R_{\rm v}\, \lambda'$, where
$V_{\rm v}$ and $R_{\rm v}$  are respectively the virial velocity  and
virial radius  of the halo,  and $J$ is the  total angular momentum of
the  halo.  The   maximum specific angular  momentum  in  the halo  is
$j_{\rm max} = j_0/(\mu-1)$.

This $M(<j)$ distribution  is considerably different from that implied
by the mass distributions in disk galaxies
\citep{b_etal01,vdb_etal01,vdb01}.  For   example,  the dark matter has
considerably  more low $j$ material than  expected  for an exponential
disk, and certain models of feedback and bulge formation do not
alleviate this problem.  It appears,  therefore, that
in order for CDM to provide a successful theory of galaxy formation, 
the   angular momentum  in proto-galactic  gas   must be
rearranged  relative to  that of the  dark  matter.

\begin{table*}
\begin{center}
\begin{tabular}{|c || c|c|c|c || c|c|c|c || c|c|c|c|}
\hline
\multicolumn{1}{|c||}{} & \multicolumn{4}{|c||}{\LCDM} & \multicolumn{4}{|c||}{\LWDM} &
\multicolumn{4}{c|}{\LWDMth} \\
\hline	
\hline
 Halo & $M_{\rm v}$ & $\mu$  & $\lambda'$  & $f_{m}$ & $M_{\rm v}$ & $\mu$  & $\lambda'$ & $f_{m}$ &  $M_{\rm v}$ & $\mu$  
& $\lambda'$ & $f_{m}$ \\

\hline 
 1 
& 9.53 & $2.13 \pm 0.55$ & 0.057  & 0.02
& 7.38  & $1.19 \pm 0.09$ & 0.043 & 0.02 
& 6.61 & $1.10 \pm 0.05$ & 0.035  & 0.04 \\

\hline
 2  
& 6.37 & $1.22 \pm 0.10$ & 0.028   & 0.00
& 5.73  & $1.16 \pm 0.07$ & 0.041  & 0.01 
& 14.7 & $1.18 \pm 0.09$ & 0.017  & 0.14\\

\hline
 3  
& 4.44 & $1.41 \pm 0.20$ & 0.023   & 0.04
& 3.83  & $1.15 \pm 0.05$ & 0.010  & 0.27 
& 3.19 & $1.10 \pm 0.04$ & 0.004  & 0.48 \\

\hline
 4  
& 5.61 & $1.06 \pm 0.02$ & 0.043  & 0.16 
& 3.76  & $1.23 \pm 0.10$ & 0.029  & 0.06 
& 2.65 & $1.07 \pm 0.03$ & 0.023  & 0.10\\

\hline
\end{tabular}
\end{center}
Table 1 -- Halo properties as obtained in each of the three simulations.  For each halo, 
listed are the values of $M_{\rm v}$ in units of $10^{13} \hMsun$, angular momentum profile shape parameter
$\mu$, along with the fit error, spin parameter, $\lambda'$, and the fraction of binned halo cell mass that
has misaligned (negative) projected angular momentum in the direction of the total halo angular momentum,
$f_m$.  
\label{tab:results}
\end{table*}

The problem of angular momentum loss and other galactic-scale problems
have   led some to   suggest  various  modifications  to  the standard
paradigm     \citep{ss_sidm,kamionkowski_liddle00}.    Among  the most
popular is     the recommendation that  WDM    be  substituted for CDM
\citep[e.g.,][]{hogan99,hogan_dalcanton00}.    This   acts to  suppress
power  relative  to CDM  on scales  below   some filtering scale $R_f$
related to the WDM particle mass.  The formation  of halos with masses
smaller than the corresponding filtering mass $M_f$ is thus delayed or
completely suppressed. The smaller number  of small-mass halos at high
redshifts ($z\gtrsim 3$) can help  to prevent the excessive transference
of angular momentum to the parent halo in cosmological  galaxy
formation simulations.

Simulations by \citet{sommer_larsen01}  indicate that  disks formed in
the  WDM  cosmology retain  a  considerably  larger  fraction of their
angular momentum than their CDM  counterparts, alleviating the angular
momentum problem discussed above and adding to  the motivation for the
WDM model.  It  is interesting therefore to  ask whether the  specific
angular momentum distribution of dark matter in the WDM is also closer
to the observed  angular momentum distribution  of baryons in observed
disks.  Implications of the  WDM  scenario have recently  been studied
extensively using cosmological simulations
\citep{colin_etal00,ar_etal01,bode_etal01,knebe_etal01,eke01}.  
It was shown
\citep{ar_etal01,bode_etal01} that WDM does help to alleviate problems
of CDM with halo  density profiles and, possibly, spatial distribution
of dwarf galaxies \citep{peebles01}. In addition,
\citet{knebe_etal01} compared specific angular momentum distribution
in CDM halos to that of the halos formed in WDM cosmology and found no
systematic difference.  However,  they studied halos  with masses much
higher than the  filtering mass  of the  simulation and neglected  the
thermal velocities of  the  WDM  particles.   The differences  in  the
specific angular momentum  distribution are especially pronounced  for
dwarf galaxies  \citep{vdb_etal01} and it  is therefore interesting to
study the angular momentum  distribution for halos below the filtering
mass.

Additional motivation for    this  project  comes from   a  desire  to
understand  the  nature  of  angular momentum   acquisition   in halos,
including the origin  of the $j$  distribution.  Such  an understanding
could provide valuable insight into the processes of galaxy formation,
e.g., by highlighting specific reasons  why (and how) the baryonic $j$
distribution  should differ     from   that  of  the    dark   matter.
Traditionally, angular momentum  in dark  halos  has been  thought  to
derive   from proto-halo   interactions  with the    large-scale tidal
field~\citep{peebles69,doro70,white84}.  Recently, however, a scenario
in    which mergers  play a  primary   role in  halo angular  momentum
acquisition     has    been        investigated\citep{vitvitska_etal01,
maller_etal01}.   Indeed, the real  situation  may be intermediate  to
these two  pictures.  Remarkably, perhaps  unfortunately, the shape of
the $M(<j)$ profile can be reasonably well  accounted for using models
based in both  frameworks.  For example,  in B01 we  presented a model
for the origin of profile (\ref{eq:fit})  focusing on angular momentum
transfer  during minor mergers, as  well  as a  model based on  linear
tidal torque  theory coupled with halo  mass accretion histories.  Van
den Bosch (2001)  has explored a  similar  model, and Maller  \& Dekel
(2001) have investigated a  detailed picture for the angular  momentum
distribution based on multiple  satellite accretion events.  Formation
of such halos in the WDM cosmology should involve very few mergers and
one might expect a  different  resulting distribution of  the specific
angular  momentum   if  mergers   are  important  in  a   halo's  spin
acquisition\citep{vitvitska_etal01, maller_dekel01}.

In  this  {\sl  Letter\/}    we use  high-resolution   dissipationless
simulations of virialized halos formed in  the CDM and WDM cosmologies
to study the  effects   of suppressed  small-scale  power and  thermal
velocities of DM  particles  on the  halo's specific  angular momentum
distribution. In  particular, we focus on  the halos with masses below
the filtering   mass of  the WDM  model.   The paper  is  organized as
follows. In  \S~\ref{sec:sims}   we  briefly  describe the   numerical
simulations used in our study and present  our analysis and results in
\S~\ref{sec:results}. Discussion  of the  results and  our conclusions
are presented in \S~\ref{sec:discussion}.

\section{Simulations}
\label{sec:sims}

To study the specific angular momentum distribution of DM halos we used 
three simulations; one of the {\LCDM} cosmology and two of
the {\LWDM} cosmology. The simulations are
described in detail in \citet{ar_etal01}; here we will briefly
summarize their main features. All simulations were run using the
Adaptive Refinement Tree code \citep{kkk97,k99} which achieves high
resolution by adaptively refining the initial uniform grid in the
regions of interest. The simulations used here were done in a
$60\hMpc$ box and we focused our analysis on the four most massive
halos in this box. The four halos were first selected from a
low-resolution $64^3$ particle run. The particles within two
virial radii of the halo centers were traced back to the initial
epoch. The initial conditions in the Lagrangian volume marked by the
particles were then reset with higher resolution using the
multiple-mass technique described in \citet{klypin_etal01}.  The final
halos studied consist entirely of the highest resolution particles,
of mass $m_p = 1.1 \times 10^{9} \hMsun$.  We used
the same power spectrum for both the {\LCDM} and {\LWDM} runs but in the
{\LWDM} simulations the modes below the filtering scale, $R_f$, were
suppressed with an exponential cutoff \citep{bbks86}:
\begin{equation}
  P_{\rm WDM}(k) = P_{\rm {\Lambda}CDM}(k)\exp\left\{-0.5\left[kR_f+(kR_f)^2\right]\right\}, 
\end{equation}
where  $P_{\rm {\Lambda}CDM}$  is  the power spectrum  of  the {\LCDM}
model. The  filtering scale defines  the characteristic filtering mass
$M_f=3.65\times 10^{14}\Omega_{\rm wdm} 
 R_f^3h^{-1}{\rm\   M_{\odot}}$ (where
$R_f$ is in $h^{-1}{\rm\ Mpc}$ and $\Omega_{\rm wdm}$ is the warm dark
matter density in units of critical density)  and is related to the mass
of         the       warm       dark          matter          particle
\citep[e.g.,][]{sommer_larsen01,bode_etal01}.   The warm dark   matter
particles should also  possess ``thermal'' velocities  with an amplitude
that is related to their  mass. In this  study we bracket the possible
effects  of  the thermal  velocities   by comparing a simulation without
thermal velocities  ({\LWDM}) and  a simulation with thermal  velocities
sixteen  times the value  expected for the  assumed WDM particle mass,
$m_W=125{\ \rm eV}$ ({\LWDMth}).  The  same set of waves was used
to set up initial conditions of both the  {\LCDM} and WDM simulations.
Therefore, the same halos form in all  the simulations which allows us
to compare   them  individually and gauge   the  effects of  the power
spectrum cutoff and thermal velocities more accurately.  

\section{Results}
\label{sec:results}


We study  four  halos in   each  simulation;  each is  identified   by
position, and numbered  1-4.  
The  studied  sample of four halos  is   too small to  draw
statistical conclusions.  However, the  identical setup of the initial
conditions  in  CDM  and  WDM  simulations has allowed  us  to  study {\em
systematic\/}  differences  between {\em  the same\/} individual halos
formed in two different cosmologies.

Halo  angular momentum  profiles  were
constructed  using the methods discussed in  B01.  Briefly, halos were
identified using   a  spherical overdensity method,   with  the virial
radius, $R_{\rm   v}$,  set  using the   standard   virial overdensity
criterion.   Once the halo is    defined,  we use  the total   angular
momentum  in each halo to assign  the $z$ direction. We then subdivide
the spherical halo volume into many spatial cells, as outlined
here using spherical coordinates about the halo center ($r$, $\theta$,
$\phi$).  Radial   shells from $r=0$ to  $R_{\rm  v}$ are defined such
that  each contains approximately the  same  number of particles.  The
number of shells is always fewer than $30$, and we demand at least 500
particles per shell.  Each radial shell is  then subdivided into three
azimuthal cells of equal volume between $\sin
\theta = 0$ and $1$, each spanning the full $2
\pi$ range  in $\phi$.  For the  halos in  this examination, each cell
contains between $500$ and $1000$ particles.   The value of $j$ in the
$z$  direction is  measured  for each cell   and  $M(<j)$ profiles are
constructed by counting the  cumulative  mass  in cells  with  angular
momentum less than $j$.

\begin{figure*}[ht]
\pspicture(0,10.0)(13.0,20.6)
\rput[tl]{0}(1.5,20.8){\epsfysize=15.5cm
\epsffile{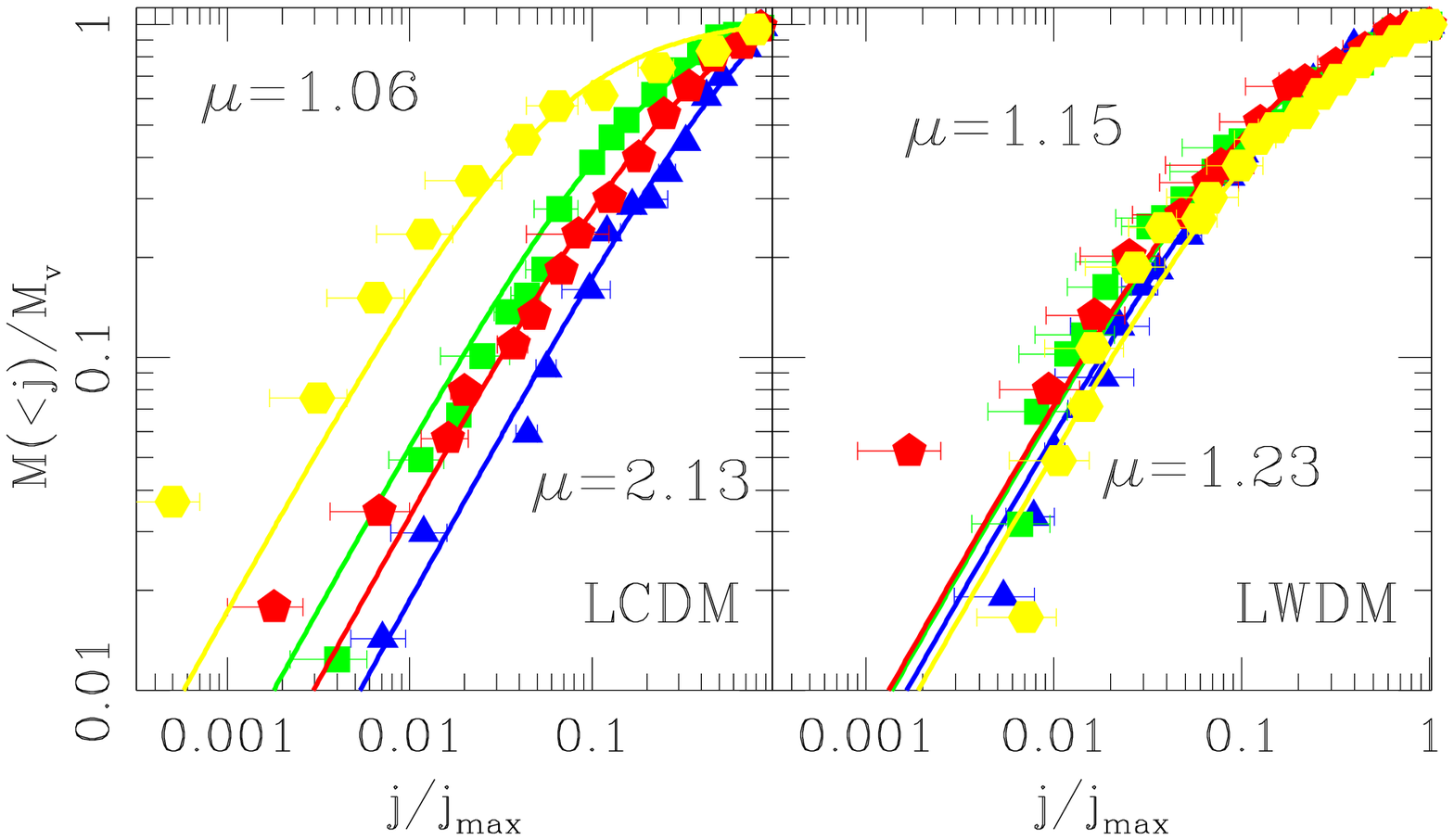}}
\rput[tl]{0}(0.0,11.2){
\begin{minipage}{18.4cm}
  \small\parindent=3.5mm {\sc Fig.}~1.--- Cumulative specific angular
momentum distributions for the {\LCDM} and {\LWDM} halos.  Filled points
show the profiles as measured in from the simulated halos and solid
lines show the best-fit profiles given by Equation 1.  Individual halos
(identified by matching positions in the two simulations) retain the
same point types in both the {\LCDM} and {\LWDM} panels.  
Although no systematic
differences are observed, the {\LWDM} halos demonstrate significantly less
scatter about the mean profile.
\end{minipage}
}
\endpspicture
\end{figure*}

Since $j$ is a projected component, it is  possible for a cell to have
a negative $j$ value.  Although this is rare (for cells containing
 $\sim 1 \%$ of the total halo  particles, as do ours), it does occur
on occasion in  CDM halos (B01), and,  as discussed below, seems to be
more  common for WDM halos.   When a cell's projected angular momentum
is negative, we remove the cell completely from the constructed $M(<j)$
profile (and do not include the  mass in any  fits).  We record the
fraction of halo mass with negative $j$ and designate it as $f_m$.
Quoted $\lambda'$ values do include the particles contained in
the negative $j$ cells
, but 
neglecting them results in changes of less than $5\%$ in $\lambda'$
in all halos except {\LWDMth} halo 3.

Table  1 lists the  angular momentum properties of all four halos as
well as their masses in each of the three simulations.  We find that
the $M(<j)$ profiles of every halo are well-fit by the universal curve given by
Equation 1.   As illustrated in Figure 1, the WDM halos show no systematic
difference  in their $j$ distributions compared to the CDM halos, 
although  there  is a dramatic
decrease in scatter about the average profile.

Note that  the  masses of the  halos   systematically decrease  as the
cosmology shifts from {\LCDM}  to  {\LWDM} and again to   
{\LWDM$_{\rm th}$}.  The  pattern holds for  all  halos except   number 2, which,  by
chance, has  just  experienced a major  merger  in  the {\LWDMth}
simulation.  The same halo is undergoing a merger  in {\LWDM}, but the
merger has not yet occurred in {\LCDM}.
A  similar  trend is   apparent in the    values of $\lambda'$,  which
decreases  systematically from   {\LCDM}  to  {\LWDM}  to {\LWDM$_{\rm
th}$}.  The lone  exception is Halo 2 in  the {\LCDM} simulation,  which,
because of its lack of a recent merger,  may not provide a fair comparison.
Similarly, the {\LWDM$_{\rm th}$} halos are significantly
more misaligned than the  {\LCDM} halos.  Because the thermal velocities
in  this model are  quite high, even at  the  time of halo turn-around
(Avila-Reese et al. 2001),    they  can influence the    measured halo
angular momentum  greatly.  The total   bulk rotation at a  given halo
radius  is  typically quite  small (i.e. $\lambda'   \sim 0.01$), so a
sizable thermal  component  can wash  out the spin  signal, leading to
significant misalignment in the measured angular momentum.

\section{Discussion and conclusions}
\label{sec:discussion}

We found no evidence for a systematic difference in  the shapes of the
specific angular momentum  distributions of WDM and CDM
halos. This  is a rather puzzling result  because there are reasons to
expect that  the shape of the  angular momentum distribution should be
sensitive    to    the       merging    history    of     the    halos
\citep{vitvitska_etal01,maller_etal01}.   A  similar   conclusion  was
reached in a recent  study by \citet{knebe_etal01}.  Their conclusion,
however,  is  not very surprising  because  it applied  to  halos with
masses  well  above  the  filtering mass.
For halos with masses below the filtering  mass, on the other hand,
the number of mergers  should be greatly suppressed, simply because
the abundance of $M<M_f$ halos is suppressed. Our results thus clearly
indicate that  minor mergers cannot play a  crucial  role in shaping
the angular momentum distribution.  This result poses a challenge for
models  striving to explain  the   angular momentum distribution as  a
result of a series of mergers.  Larger-scale phenomena or
processes associated with halo virialization may be the major
contributors.

We found evidence that WDM halos possess
systematically    smaller  spins than    their   counterparts  in  the
{$\Lambda$CDM} model, although this is based on only three halos.
The forth halo does not follow this trend, however the comparison
is somewhat biased because the WDM halo is undergoing a merger
~\footnote{Interestingly, the merger does not drastically affect the 
shape of the angular
momentum profile.}.
This result is in agreement with findings of
\citet{knebe_etal01} who reported  systematically lower average spins
for  halos  in the  WDM models   with  higher filtering  mass  using a
statistical sample  of halos (see their Fig.~15).  We showed  that the
addition of initial thermal velocities of the WDM particles results in
even lower spins of  the WDM halos  as well as a significantly  larger
misaligned (negative $j$)   mass   fraction.  This is   a  potentially
significant    result  because   previous   galaxy   formation studies
\citep[e.g.,][]{mmw98} showed   that  the angular momentum  of halos
in CDM  models is barely sufficient  to produce disks with
realistic sizes. Although it is not clear why $M<M_f$ WDM halos 
should have smaller spins than their CDM counterparts,  in  the
context of merger-driven spin-evolution models
\citep{vitvitska_etal01,maller_etal01} it might be expected because
they accrete most of  their mass quiescently.
A decrease in spin   during periods of quiescent  accretion
 is also observed in cosmological simulations
\citep{vitvitska_etal01}.

Recently,  \citet{sommer_larsen01} performed gasdynamic simulations of
galaxy formation in   CDM and WDM  models  and argued
that the WDM    model can help  to resolve the
angular  momentum  problem.  However,   this
conclusion was true only for simulations in which disk galaxies formed
in halos with mass  $>M_f$,  while for halos  of mass  $M\lesssim M_f$
they  found  no increase  in the  specific  angular  momentums of  the
simulated     disks  compared to   the      CDM  simulations. If,   as
\citet{sommer_larsen01} argued, the gas in  the WDM model retains more
of   its  initial angular  momentum  due   to  the lower abundance  of
small-mass halos  (and hence less  efficient cooling  of gas  at early
epochs),  this should also  be true for  the  halos with $M<M_f$.  The
relatively small specific angular momentum for the $M<M_f$ should then
be due   to the lower  overall  spin  of the   gas   and DM, which  is
consistent with the results discussed above.

Unless there is a mechanism which allows  baryons to acquire more
angular momentum than the  dark matter, this result  may be problematic
for WDM models. Indeed, if, following \citet{sommer_larsen01}, one
tunes the filtering mass  to be $\sim 10^{10}-10^{11}{\ M_{\odot}}$ to
alleviate the angular momentum problem for  massive disk galaxies, the
dwarf galaxy disks  will then be forming  in the lower-spin  $M\ll M_f$
halos. The WDM  dwarf  disks should, therefore, possess  smaller spins
than disks formed in comparable mass halos in the CDM models,
which are already uncomfortably low  compared  to  the  spins  of   
the  observed  dwarf  galaxy  disks
\citep{vdb_etal01}. In   addition,  both the CDM   and  WDM halos have
similarly discrepant $M(<j)$ profiles
compared those of dwarf disks \citep{vdb_etal01}.

In conclusion, the results presented  in this {\em Letter\/} show that
the change from  cold to warm  dark  matter does not produce  the dark
matter  halos  with higher angular   momenta or  with  more  desirable
specific angular momentum distributions. For WDM disks forming in halos
with  masses  smaller than  the  filtering  mass the angular  momentum
problem might  actually be worse  than for the  disks  forming in CDM
halos of  similar  mass. Our  results therefore  fail to provide  an
additional   motivation for the   WDM  scenario, while highlighting  a
possibly severe problem at dwarf galaxy scales.

\acknowledgements 
J.S.B. aknowledges funding from NSF grant AST-9802568.
A.V.K. was supported by NASA through Hubble Fellowship grant from the
Space Telescope Science Institute, which is operated by the
Association of Universities for Research in Astronomy, Inc., under
NASA contract NAS5-26555.
Simulations were performed at the Direcci\'on General de Servicios 
de C\'omputo Acad\'emico, UNAM, using an Origin-2000 computer.

\bibliographystyle{apj}

\bibliography{jwdm}

\end{document}